\documentclass[galaxies,article,accept,oneauthor,pdftex,10pt,a4paper]{Definitions/mdpi}
\firstpage{1}
\articlenumber{96}
\pubvolume{6}
\issuenum{3}
\pubyear{2018}
\copyrightyear{2018}
\history{Received: 24 July 2018 ; Accepted: 31 August 2018; Published: 6 September 2018}
\updates{yes}

\Title{The Morphology of the Outflow in the Grazing Envelope Evolution}

\Author{Sagiv Shiber\href{https://orcid.org/0000-0001-6107-0887}{\orcidicon}}

\AuthorNames{Sagiv Shiber}

\address[1]{
Physics Department, Technion---Israel Institute of Technology, Technion City, Haifa 3200003, Israel; shiber@campus.technion.ac.il\\}

\corres{Correspondence: shiber@campus.technion.ac.il}

\abstract{
We study the grazing envelope evolution (GEE), where a secondary star, which orbits the surface of a giant star, accretes mass from the giant envelope and launches jets.
We conduct simulations of the GEE with different half-opening angles and velocities, and simulate
the onset phase and the spiralling-in phase. We discuss the resulting envelope structure and the outflow geometry.
We find in the simulations of the onset phase with narrow jets that a large fraction of the ejected mass outflows along the polar directions.
The mass ejected at these directions has the fastest velocity and the highest angular momentum magnitude.
In the simulations of the spiralling-in phase, a large fraction of the ejected mass concentrates around the orbital plane.
According to our findings, the~outflow with the highest velocity is closer to the poles as we launch narrower jets.
The~outflow has a toroidal shape accompanied by two faster rings, one ring at each side of the equatorial plane.
The interaction of the jets with the giant envelope causes these outflow structures, as we do not include in our simulations the secondary star gravity and the envelope self-gravity.
}


\begin{document}

\section{Introduction}
Numerical simulations of the common envelope evolution (CEE) have been performed throughout the last thirty years \cite{LivioSoker1988, RasioLivio1996, Sandquistetal1998, RickerTaam2012, Passyetal2012, Ohlmannetal2016, NandezIvanova2016, Iaconietal2018}.
The simulations attempt to answer fundamental questions regarding the CEE and aim to produce the observed close binary systems where at least one star is a white \mbox{dwarf \cite{Ivanovaetal2013, DeMarcoIzzard2017}}. Due to the lack of symmetry axis, these simulations are three-dimensional and they usually include only the gravitational orbital energy of the in-spiralling binary.

Table \ref{table:CESimulations} compares the results of several numerical simulations of the CEE. The second row lists the minimum final separation achieved between all the simulations presented in the same paper. The third row lists the maximal fraction of the envelope mass that was ejected from the system in each paper. Almost all simulations failed to achieve the envelope ejection and the observed small final separation. The exception is \cite{NandezIvanova2016}, where they included recombination energy as another source of energy. However, the role of recombination in common envelope removal is still in debate \cite{Claytonetal2017, Ivanova2018, Sabachetal2017, Gricheneretal2018}.

Another extra energy source that has been suggested to contribute to the envelope ejection comes from the accretion of envelope mass on to the more compact secondary star. In particular, when the secondary star accretes mass through an accretion disk and launches jets that interact with the envelope of the giant star. This possibility was first discussed in \cite{ArmitageLivio2000,Chevalier2012} for a neutron star, and then for other companions \cite{Soker2004, MorenoMendezetal2017, Camaraetal2018}.
The jets themselves remove mass, angular momentum, and energy from the vicinity of the secondary star, and by that allow further accretion \cite{Shiberetal2016, Chamandyetal2018}.

\begin{table}[H]
\caption{Comparison between results from common envelope evolution simulations. $a_{f}^{\rm min}$
denotes the minimal final separation achieved in each paper and $f_{\rm env}^{\rm max}$ denotes the maximal fraction of the envelope ejected in each paper. The fourth row states the numerical technique used in each paper, i.e., grid code or smoothed-particle hydrodynamics (SPH) code.  }
\centering
\begin{tabular}{cccccccc}
\toprule
\textbf{References}	& \textbf{\cite{RasioLivio1996}}	& \textbf{\cite{Sandquistetal1998}} & \textbf{\cite{RickerTaam2012}} & \textbf{\cite{Passyetal2012}}	& \textbf{\cite{Ohlmannetal2016}}	& \textbf{\cite{NandezIvanova2016}} & \textbf{\cite{Iaconietal2018}}\\
\midrule
$a_{f}^{\rm min}  \left(R_{\odot}\right)$ 		& 4.3			& 4.4		& 8.6			& 5.9		& 4.2			& 0.43		& 16\\
$f_{\rm env}^{\rm max}$		& 0.1			& 0.31		& 0.26			& 0.15		& 0.06			& 1		& 0.16\\
Grid/SPH		& SPH			& Grid		& Grid			& Both
& Moving Mesh			& SPH		& Both\\
\bottomrule
\label{table:CESimulations}
\end{tabular}
\end{table}

\vspace{-8pt}
Observational evidence supports mass accretion by the secondary star. For example, the observed chemically polluted secondary star in the Necklace nebula \cite{Miszalskietal2013}. In several other planetary nebulae (PNe), an inflated secondary main sequence (MS) star is found in a close orbit with the central star~\cite{Jonesetal2015}. The inflation is best explained by a recent accretion phase where the accretion rate needs to be high. Observations of Fleming 1 (PN G290.5+07.9) indicate that the formation of the jets precedes the ejection of the central nebula \cite{Boffinetal2012}. Accretion on to the secondary star, while the binary separation was still large, and, during a relatively long-lived period, caused the launching of the jets. Only later, a faster binary interaction created the central nebula.

In 2015, it was proposed that jets launched from the secondary star can prevent in some cases the system from entering the CEE \cite{Soker2015}. The result is a new evolutionary phase that was termed the grazing envelope evolution (GEE).
The system initial setup is of a nearly contacting binary where a secondary star already has an accretion disk as it approaches a giant star. The accretion disk launches two opposite jets that remove envelope mass and cause the giant radius to shrink simultaneously with the orbital separation.
If jets are inefficient at removing mass, the system might enter a CEE.

First simulations of the GEE focused on simulating the initial evolution of the GEE \citep{Shiberetal2017}. A $0.5~ M_{\odot}$ MS star was placed on a circular orbit close to the surface of a $ 3.4~ M_{\odot}$
asymptotic giant branch (AGB) star. The simulations aimed to test the influence of the jets on the giant envelope, hence they did not include the secondary stellar gravity and the self-gravity of the envelope. The secondary star launched jets perpendicular to the orbital plane with a half-opening angle of 30 degrees while circling the giant. The jets were launched with  a velocity of approximately the escape velocity from a low mass MS star, at a velocity of $700\; {\rm km\; s^{-1}}$. The mass injected in the jets was taken to be between one to five percent of the Bondi–Hoyle–Lyttleton (BHL) accretion rate \cite{Bondi1952} on the giant surface. After eight orbital periods, the interaction of the jets with the envelope ejected between $0.05~ M_{\odot}$ to $0.25~ M_{\odot}$ of the envelope gas. Approximately 90 percent of the ejected gas was unbound.

In a later, more extended study the companion was set to spiral in to the envelope while it launches jets \cite{ShiberSoker2018}. These simulations included different opening angles and velocities of the jets. The companion was set to move from the giant surface to two-thirds of the giant radius in a constant inward radial velocity while circling the giant in a tangential Keplerian velocity.  The companion spirals-in in a time of three initial orbital periods, in a way that resembles orbits of some other CEE simulations. All the rest of the initial conditions (the giant profile, the secondary star mass, etc.) remained identical to the initial conditions of the onset of the GEE simulations.
The results from this study indicate that, regardless of the different jets properties, the jets succeed at expelling mass from the spiraling-in secondary star vicinity and postpone the CEE.

Here, we focus on the morphological structure of nebulae formed by binary systems that have gone through the GEE. This is important because, in this way, we can differentiate between CEE and GEE while analyzing observations.
{{{The morphological structure we obtain in the numerical simulations can critically be compared to observations. Observers can link specific systems to this evolutionary phase which then could lead to predictions regarding the further evolution of such systems.}}} 
In Section \ref{sec:results}, we describe the results and, in Section \ref{sec:discussion}, we discuss the results and summarize.

\section{Results}
\label{sec:results}

The results from the simulations, both from the onset of the GEE and from the spiralling-in orbits, show several structures. We present new simulations of jets with a half-opening angle that equals 15~degrees, i.e., narrow jets.
We describe the morphologies of the descendant nebulae from the onset of the GEE in Section \ref{subsec:onset} and from the spiralling-in phase in Section \ref{subsec:spiralin}.


\subsection{The Onset of the GEE}
\label{subsec:onset}

\begin{figure}[H]
\centering
\includegraphics[trim= 0.0cm 0.7cm 0.0cm 0.4cm,clip=true,width=0.48\columnwidth]{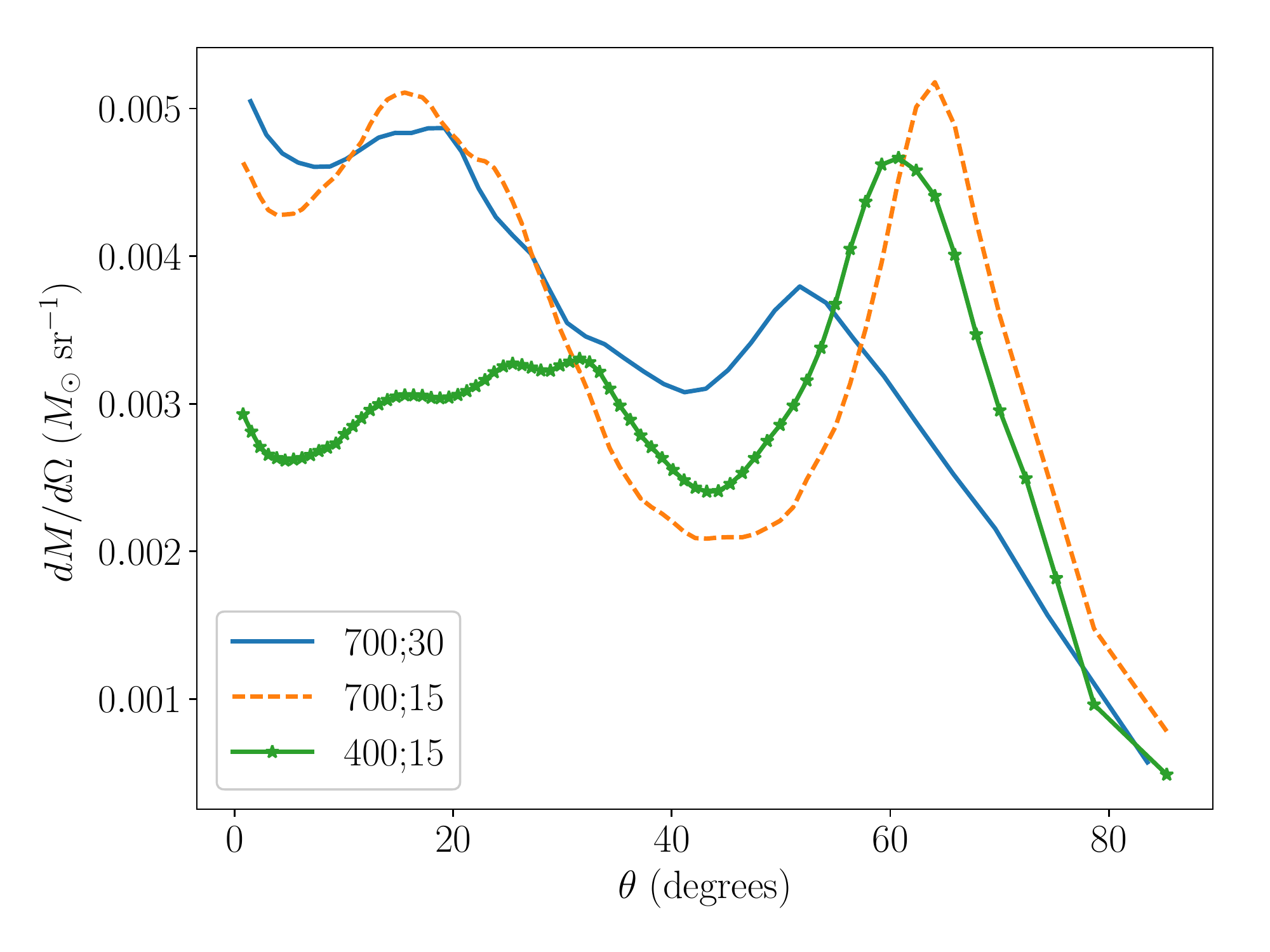}
\includegraphics[trim= 0.0cm 0.7cm 0.0cm 0.4cm,clip=true,width=0.48\columnwidth]{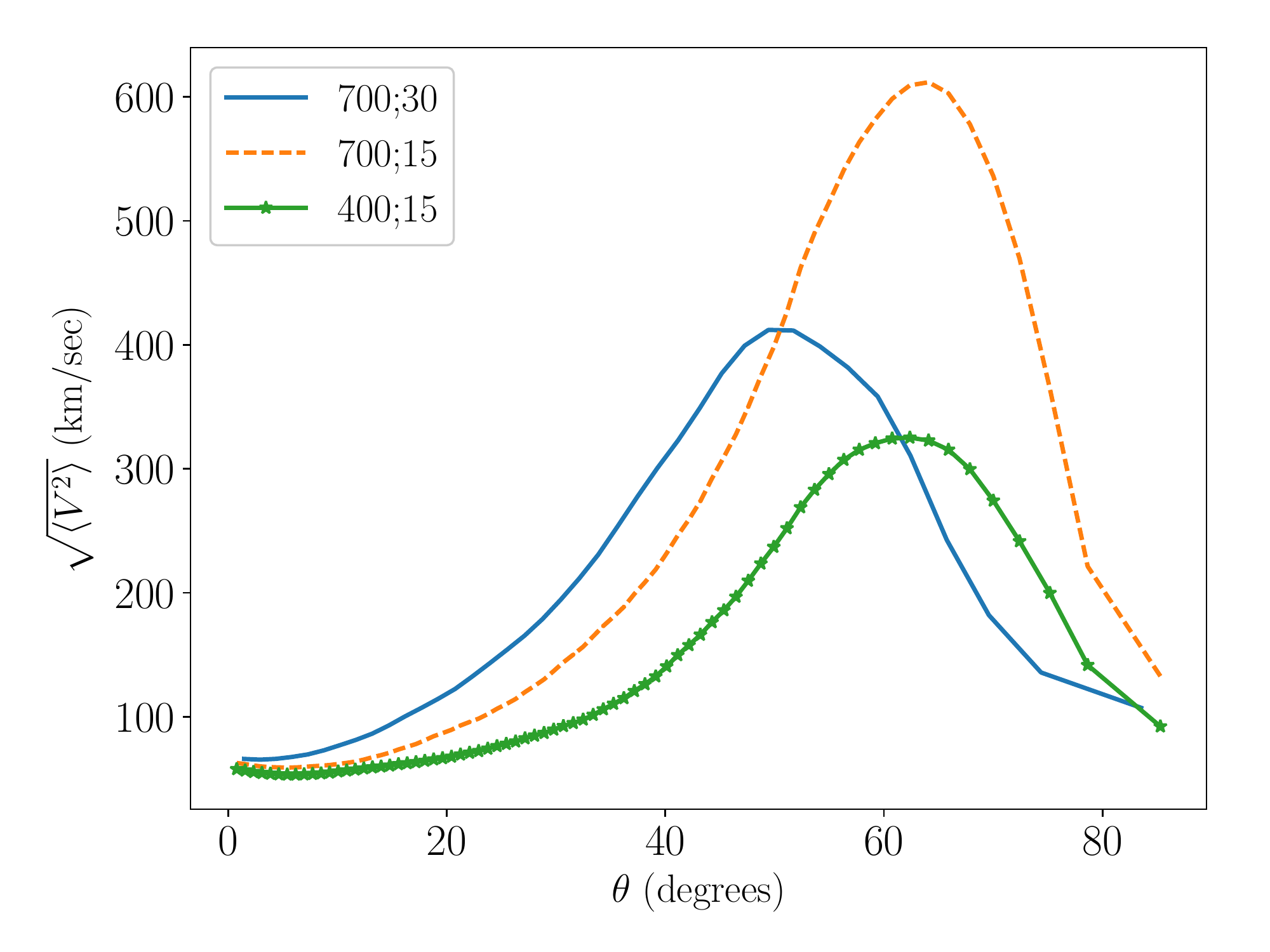}
\caption{\textbf{Left panel}: The mass lost from the grid per unit solid angle as a function of the angle from the equatorial plane, for three different simulations of the onset of the grazing envelope evolution. {{{The secondary star launches jets while it circles the asymptotic giant branch star on its equator in a
Keplerian orbit with a period of 198.4 days. The simulations stop after $1620$ days.}}} The three first digits in the legend represent the jets velocity in kilometres per second while the two last digits represent the jets half-opening angle in degrees. The equator is at $\theta=0$ and the poles are at $\theta=90$. Simulation 700;~30 is from \cite{Shiberetal2017}. The amount of mass lost is that from the two hemispheres combined. In all three cases, the jets inject mass at a rate of $\dot{M}_{j} = 0.001\; M_{\odot} \;{\rm year^{-1}}$. \textbf{Right panel}: The average speed of the outflow as a function of the angle from the equatorial plane of these three simulations.}
\label{fig:onsetm}
\end{figure}

At the onset of the GEE, the secondary star orbits in a Keplerian circular motion on the surface of the giant star while launching jets. We simulated two new cases of jets with a half-opening angle of 15~degrees, {{{ which are different only in their jets velocity}}}. 
We compare these two new simulations to a simulation with wider jets of 30 degrees half-opening angle.
In all three cases, the jets inject  {mass at a rate} of $\dot{M}_{j} = 0.001\; M_{\odot} \;{\rm year^{-1}}$.
Figure \ref{fig:onsetm} shows the mass ejected from the system and the outflow average speed as a function of angle from the equatorial plane. The legends show the properties of the simulations. The three first digits at each row represent the jet velocity in kilometres per second while the two last digits represent the jet's half-opening angle in degrees.
We find that narrower jets cause higher mass ejection at mid-latitude.
In fact, we obtained a bipolar nebula. Moreover, according to \cite{Franketal2018}, the fast post-AGB wind, which will later interact with the ejected envelope, will enhance the density contrast in the descendant planetary nebula (PN).

\begin{figure}[H]
\centering
\includegraphics[trim= 0.0cm 0.7cm 0.0cm 0.4cm,clip=true,width=0.48\columnwidth]{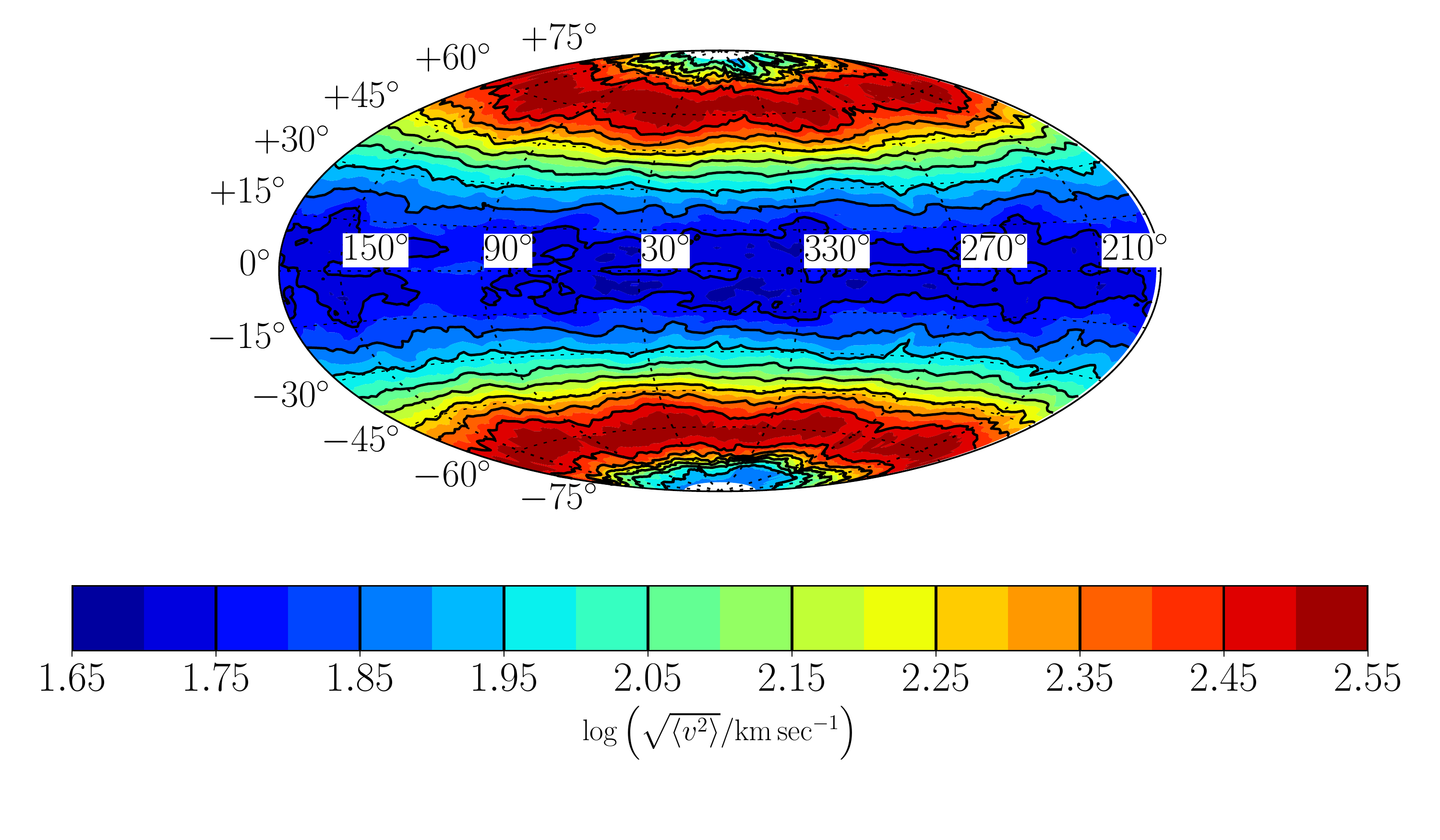}
\includegraphics[trim= 0.0cm 0.7cm 0.0cm 0.4cm,clip=true,width=0.48\columnwidth]{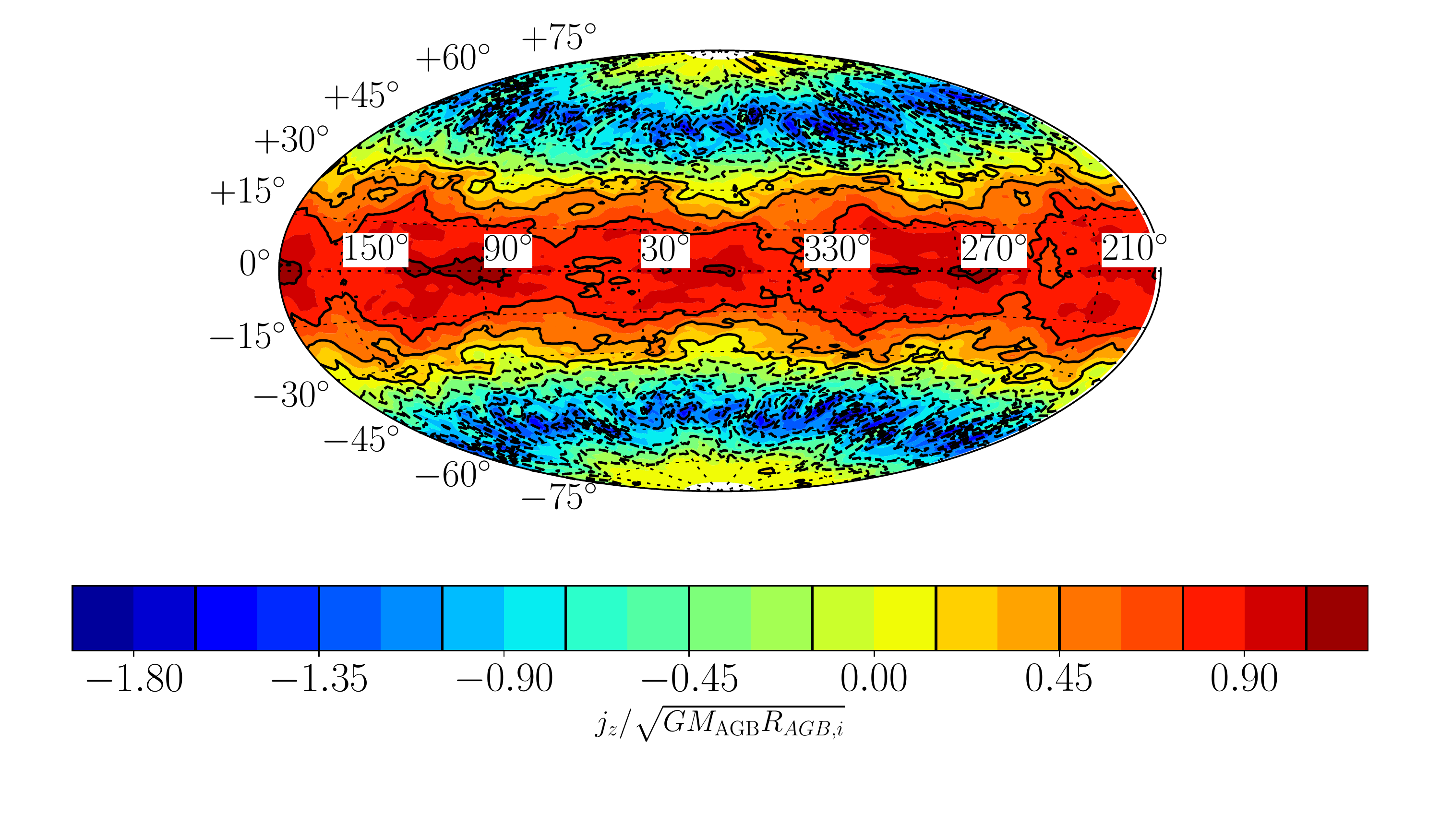}\vspace{-6pt}
\caption{\textbf{Left panel}: The average speed of the outflow at a spherical shell of radius 4 AU for a simulation of the onset of the grazing envelope evolution with jet velocity of $400 {\rm\; km\; s^{-1}}$ and jet's half-opening angle of $15^{\circ}$. \textbf{Right panel}: The average specific angular momentum in the $z$-direction (perpendicular to the orbital plane) in the same sphere of the same simulation. The average specific angular momentum is divided by the initial specific orbital angular momentum of the binary system. Zero latitude is the equatorial (orbital) plane and zero longitude (at the center) is the initial location of the companion. The companion is moving towards higher angles, namely from the right to the left.
}
\label{fig:onsetoutflowmaps}
\end{figure}

To further illustrate the effect of narrow jets, we focus on the simulation with jets velocity of $400 \;{\rm km\; s^{-1}}$.
We sum over the mass and the kinetic energy that leave a sphere at the edge of the grid and derive the average speed of the outflow as function of the latitude and longitude. We also calculated the average specific angular momentum of the outflowing gas.
Figure \ref{fig:onsetoutflowmaps} shows maps of the average speed and of the average specific angular momentum in the $z$-direction (perpendicular to the orbital plane) of the outflow. The average specific angular momentum is divided by the initial specific orbital angular momentum of the binary system. The maximum in the outflow average velocity (the red area in the left panel of Figure \ref{fig:onsetoutflowmaps}) coincides with the angle of major mass ejection and with the maximum in negative average specific angular momentum (the blue area in the right panel of Figure~\ref{fig:onsetoutflowmaps}).

Most of the outflowing gas concentrates in two expanding rings, one at each side of the orbital plane at latitudes of of $\pm 70^\circ$, which contain the fastest outflowing gas. The magnitude of the angular momentum in these rings is the largest, {\it but in an opposite direction to the orbital angular momentum of the binary system}.
The rings spin in an opposite direction due to the bending backwards of the jets by the dense envelope, an effect we have already seen and discussed in earlier studies \cite{Shiberetal2017, ShiberSoker2018}. The total angular momentum that leaves the computational domain is in an opposite direction to the orbital angular momentum of the binary system. As a result, the remaining bound envelope material mildly spins~up.

{{{{We expect that the descendant nebula created by this outflow will exhibit a mirror symmetry. A~mirror symmetry has been observed in many bipolar nebulae, such as nebula M2-9 \cite{LivioSoker2001}.
The remnant of supernova (SN) 1987A shows a mirror symmetry with two quite symmetric outer rings that expand with a radial velocity of approximately $20\;{\rm km\;s^{-1}}$ \cite{Burrowsetal1995}, a slower velocity than the outflow velocity we obtain. This difference can be explained by the fact that we do not include radiative cooling in our~simulations.}}}}

{{{{The results from the onset phase of the GEE can be compared to a class of post-AGB stars that have a companion with an intermediate orbital period, typically between one hundred to several thousands of days \cite{Manicketal2017}. Their orbital periods lay in the gap of the traditional bimodal distribution of post-AGB binaries \cite{Nieetal2012}.
Several cases of jets were observed within these systems with velocities that indicate the presence of a MS companion \cite{vanWinckel2017}. Moreover, in one such system, BD+$46^{\circ}442$, the half-opening angle was measured, assuming a certain inclination, to be roughly $60^{\circ}$ \cite{Bollenetal2017}. Doppler tomography of this system reveals that the jet launching happens around the companion. Based on our results, we predict that post-AGB intermediate binaries (post-AGBIB) that launch narrow jets will produce bipolar nebulae. A known example for such a system is the red rectangle, HD 44179 \cite{vanWinckel2014}. }}}}

\subsection{GEE with Spiralling-in Orbits}
\label{subsec:spiralin}

\begin{figure}[H]
\centering
\includegraphics[trim= 0.0cm 0.7cm 0.0cm 0.4cm,clip=true,width=0.48\columnwidth]{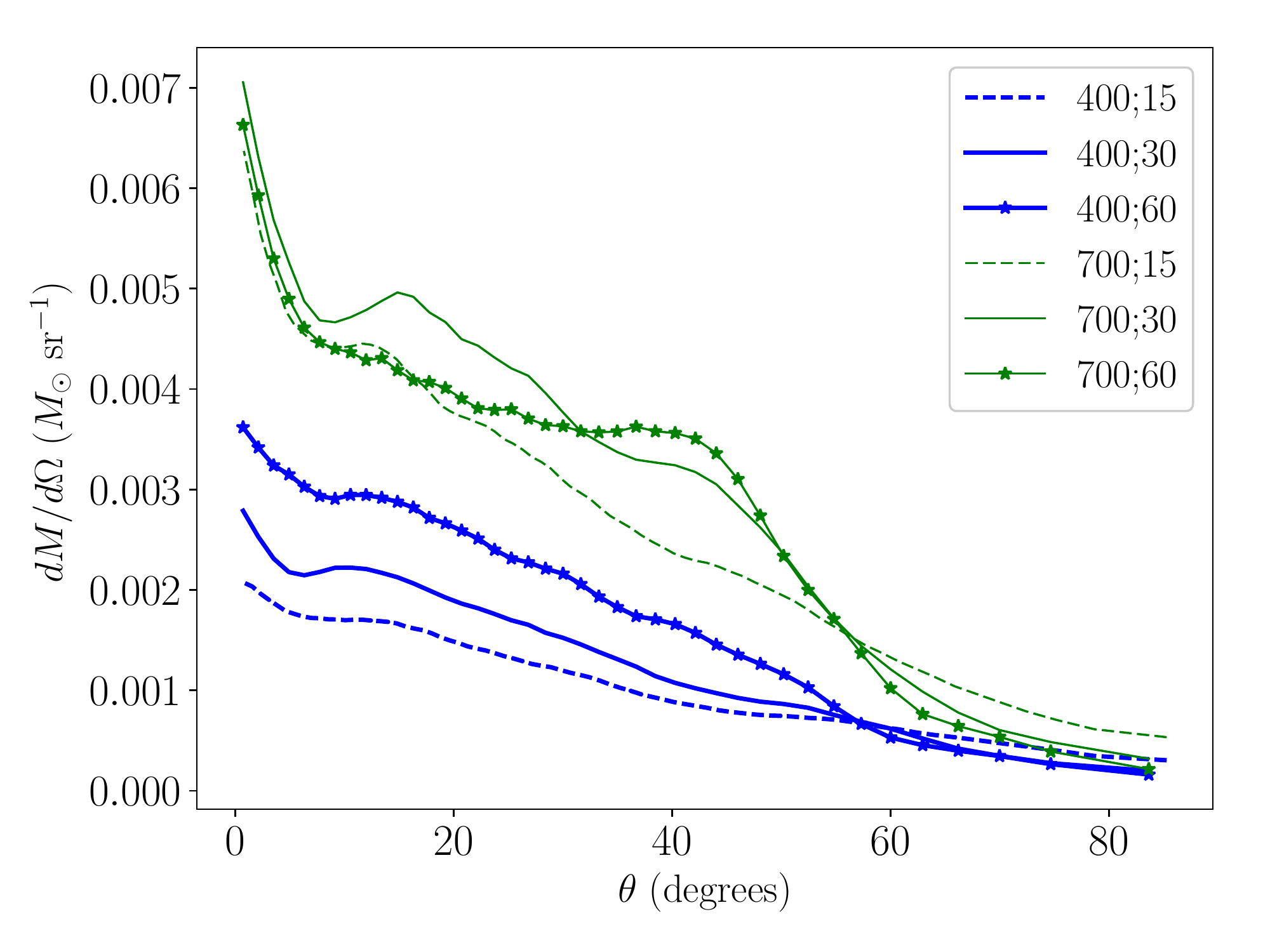}
\includegraphics[trim= 0.0cm 0.7cm 0.0cm 0.4cm,clip=true,width=0.48\columnwidth]{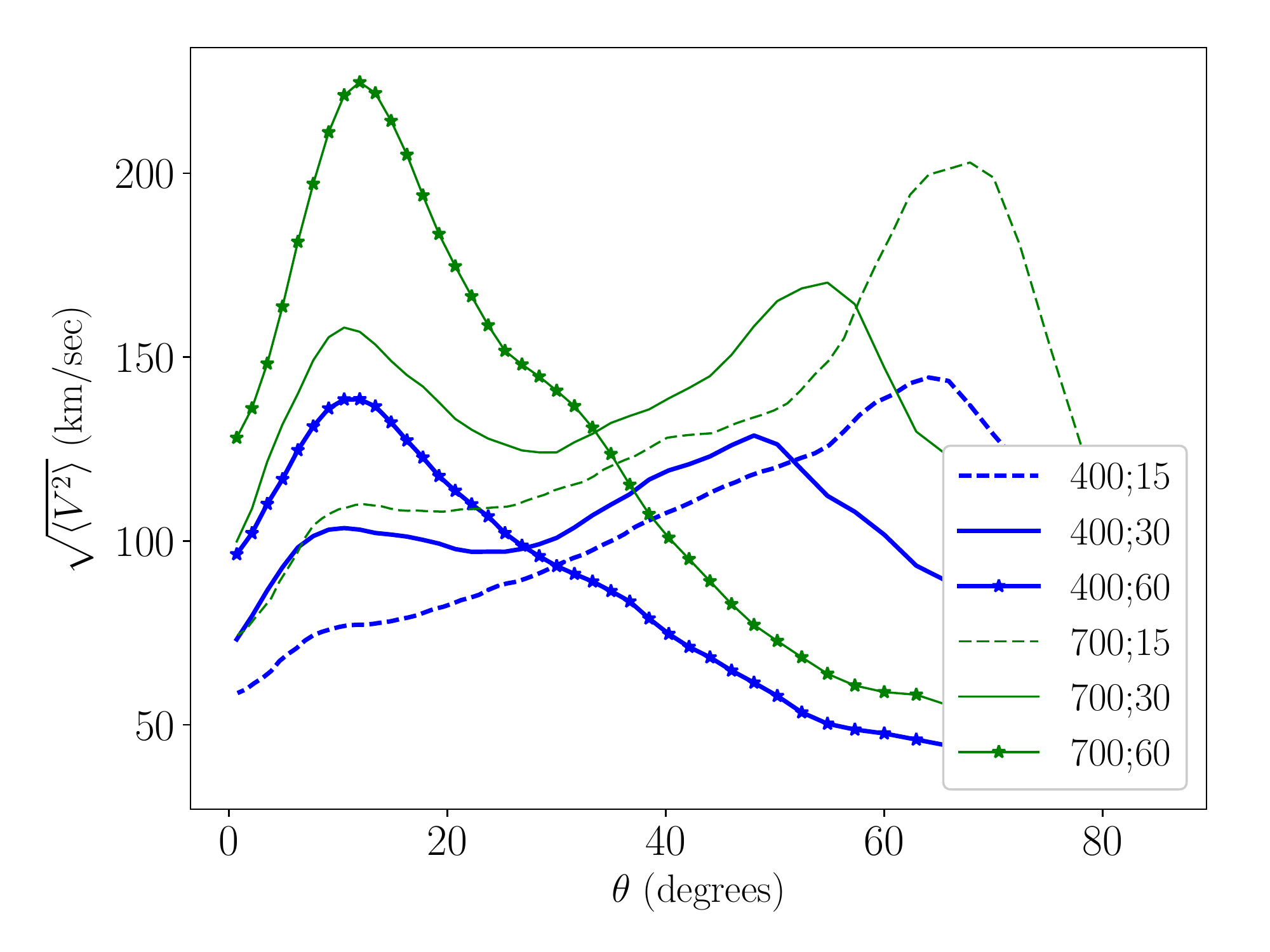}
\caption{ {{{{\textbf{Left panel}: The mass lost from the grid per unit solid angle as a function of the angle from the equatorial plane, for six different simulations with spiralling-in orbits. The secondary star launches jets while it spirals-in the asymptotic giant branch star envelope on its equator with a tangential Keplerian velocity and with a constant inward radial velocity. The initial orbital period equals to 198.4 days. The simulations stop after $595$ days. The three first digits in the legend represent the jets velocity in kilometres per second while the two last digits represent the jets half-opening angle in degrees. The equator is at $\theta=0$ and the poles are at $\theta=90$. Simulations: 400;30, 400:60, 700;30, 700;60 are from \cite{ShiberSoker2018}, while simulations 400;15 and 700;15 are new. The amount of mass lost is that from the two hemispheres combined. In all six cases the jets inject mass at a rate of $\dot{M}_{j} = 0.001\; M_{\odot} \;{\rm year^{-1}}$. \textbf{Right~panel}: The average speed of the outflow as a function of the angle from the equatorial plane of these six simulations.}}}} }
\label{fig:spiral_vrms_latitude}
\end{figure}

The simulations with spiralling-in orbits show mass ejection mainly at low latitudes (see left panel in Figure \ref{fig:spiral_vrms_latitude}), i.e., an equatorial dense outflow. The reason is the deflection of the jets towards the equatorial plane when the companion orbits inside the dense envelope. Conventional CEE simulations, which do not include jets, also lead to a dense equatorial outflow. The gravitational interaction of the envelope with the secondary star causes the equatorial outflow. We do not include the secondary star gravity in our simulations.
We expect that the combined effect of the jets with the secondary gravity will produce a pre-PN and later a PN with a massive torus.

The right panel of Figure \ref{fig:spiral_vrms_latitude} presents the average speed of the outflow as a function of the latitude in the simulations of the GEE with spiralling-in orbits.
The two simulations with the narrow jets are new, while the four simulations with wider jets are from \cite{ShiberSoker2018}. We see that narrow jets shift the highest velocity outflow towards the poles.
Simulations of jets with a half-opening angle of 60 degrees have a definite peak in the outflow speed at low latitude, while simulations with narrow jets (dashed lines in the right panel of Figure \ref{fig:spiral_vrms_latitude}) have a definite peak in the outflow speed close to the poles.
The escape velocity from the giant star at the boundary of the grid (a distance of 4 AU) equals to approximately $40 {\rm \; km \; s^{-1}}$, thus, as seen in Figure \ref{fig:spiral_vrms_latitude}, a large fraction of the ejected gas is unbound.

\begin{figure}[H]
\centering
\includegraphics[trim= 0.0cm 1.2cm 0.0cm 0.4cm,clip=true,width=0.75\columnwidth]{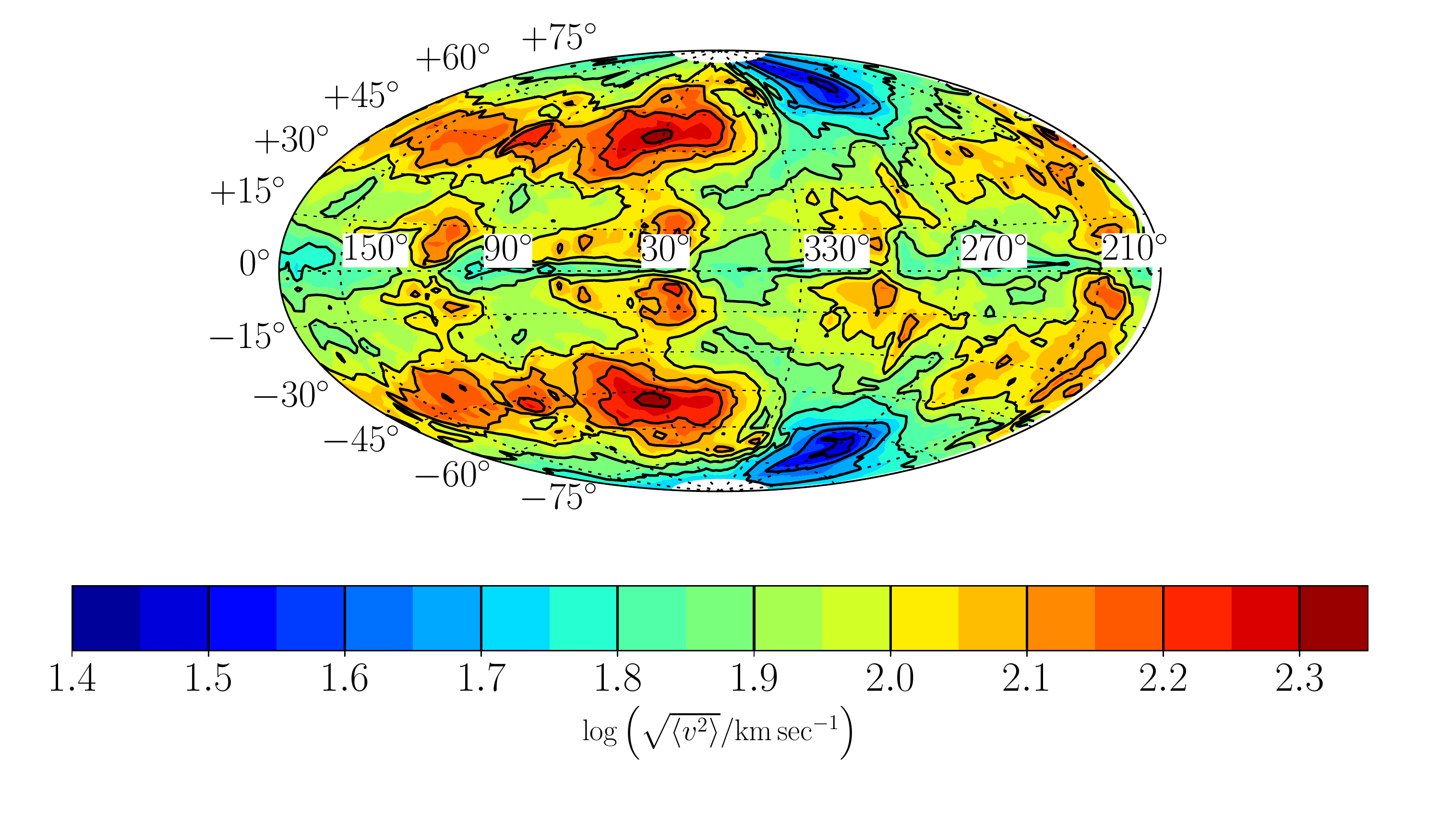}\vspace{-4pt}
\caption{The average speed of the outflow at a spherical shell of radius 4 AU for a simulation with spiralling-in orbit with jets velocity of $400 {\rm\; km\; s^{-1}}$ and jets half-opening angle of $30^{\circ}$. Zero latitude is the equatorial (orbital) plane and zero longitude (at the center) is the initial location of the companion. The companion is moving towards higher angles, namely from the right to the left.}
\label{fig:GEE40030vrms}
\end{figure}

We plot in Figure \ref{fig:GEE40030vrms} the average speed of the outflow as a function of the latitude and longitude for the case of jets with a half-opening angle of 30 degrees and jets velocity of $400 {\rm \; km\;s^{-1}  }$.

While there are sporadic peaks near the equatorial plane, there are extended peaks at around $\pm 45^\circ$ latitude, which lead to the formation of arcs of high velocity outflow, one at each hemisphere. The~sporadic peaks in the outflow average speed near the equatorial plane together with the clumpy mass ejection found near the equatorial plane \cite{ShiberSoker2018} can create the knotty waists observed in the Necklace nebula and in some other similar PNe, such as Hen 2-161 \cite{Miszalskietal2013, Jonesetal2015}. We note that the equatorial plane in SN 1987A also shows knotty features.

To conclude, the morphology of the descendant nebula in the spiralling-in simulations is of a massive torus surrounding the central binary. Fast rings or arcs of material expand below and above the torus.
The size of the rings, their number, and even their distance at a given time are the imprints of the jets interaction with the giant envelope.

\section{Discussion}
\label{sec:discussion}

We examined the outflow morphology of the GEE as a result of jets that a secondary
star launches when it accretes mass from the envelope of a giant star. We conducted hydrodynamical
simulations of two phases, the onset of the GEE, when the secondary star orbits the giant star in a circular Keplerian motion at the giant surface, namely, it grazes the envelope from outside, and the spiralling-in phase, when the secondary star grazes the envelope from inside as it spirals-in.

We presented simulations with jets of different half-opening angles and velocities. The simulations with narrow jets, of a half opening angle of 15 degrees, are presented for the first time for the GEE.
Simulations of the onset of the GEE with narrow jets show massive ejection near the poles (left
panel of Figure \ref{fig:onsetm}).
Jets with a larger half-opening angle of 30 degrees during the onset phase, and all
of the jets that we simulated in the spiralling-in phase, interact with the envelope and create a dense equatorial
outflow. This will later form a nebula with a dense expanding torus, {{{{which might have knotty structures}}}}. Rings (right panel of Figure \ref{fig:spiral_vrms_latitude}) or arcs (Figure \ref{fig:GEE40030vrms}) of fast outflow appear at mid-latitudes below and above the central torus (left panel of
Figure \ref{fig:spiral_vrms_latitude}).

{{{{ The structure of a knotty equatorial torus and a pair of faster rings, one at each side of the equatorial plane and at mid latitudes, is similar to the structure of the three rings of SN 1987A \cite{Burrowsetal1995}. We~suggest that the three rings of SN 1987A were formed by a GEE, with a companion that later entered the envelope and collided with the core of the progenitor of SN 1987A. The three rings expand in a relatively slow speed, much slower than the velocities we obtain. It is possible that the inclusion of radiative cooling into our simulations will yield lower outflow velocities.}}}}

{{{{We postulate that the observed post-AGBIBs are systems that experienced in the past or are experiencing at present the GEE. Our simulated orbital periods are consistent with the short end of the post-AGBIBs orbital periods. Indeed, a recent study used the MESA BINARY code to show that systems experiencing a significant mass ejection by jets can avoid the CEE \cite{Abu-Backeretal2018}. }}}}

{{{{At this stage, we conducted three-dimensional hydrodynamical simulations only of the onset phase of the GEE and of the spiralling-in phase of the GEE.}}}}
After the onset of the GEE, the system can either continue in the GEE or in the CEE.
The exact conditions that determine the evolutionary route are still unclear.
To test the transition from the onset phase of the GEE to a long-lasting GEE
or to a CEE, simulations of the GEE that include the secondary gravity and the envelope self-gravity
should be performed. This will be the subject of a forthcoming study.

\vspace{6pt}

\funding{This research received funding from the Israel Science Foundation.}

\acknowledgments{{{{{I thank the referees for useful comments that helped improve the paper. I thank Noam Soker for useful discussions.}}}}
This work was supported by the Cy-Tera Project, which is co-funded by the European Regional Development Fund and the Republic of Cyprus through the Research Promotion Foundation.
}

\conflictsofinterest{The author declares no conflict of interest.}

\reftitle{References}

\end{document}